\documentclass[sigplan,nonacm]{acmart}
\pdfoutput=1
\usepackage[]{hyperref}
\usepackage{makecell}
\usepackage{caption}
\usepackage{pifont}
\usepackage{xspace}
\usepackage{enumitem}
\usepackage[most]{tcolorbox}
\usepackage{algorithm}
\usepackage[noend]{algpseudocode}

\newcommand{\sysname}{\texttt{DDiT}\xspace}

\usepackage{tikz}

\begin{document}

\title{\sysname: Dynamic Resource Allocation for Diffusion Transformer Model Serving}

\pagestyle{plain}

\renewcommand{\thefootnote}{\fnsymbol{footnote}}

 \author{
 \rm
 Heyang Huang$^{*}$ \textsuperscript{1,2}
 Cunchen Hu$^{*}$ \textsuperscript{1,2},
 Jiaqi Zhu\textsuperscript{1,2},
 Ziyuan Gao\textsuperscript{1,2},
 Liangliang Xu\textsuperscript{3},
 Yizhou Shan\textsuperscript{4},
 Yungang Bao\textsuperscript{1,2},
 Sun Ninghui\textsuperscript{1,2},
 Tianwei Zhang\textsuperscript{5},
 Sa Wang\textsuperscript{1,2}
 \break \\
 \textsuperscript{1}\textit{University of Chinese Academy of Sciences}
 \textsuperscript{2}\textit{State Key Lab of Processors, Institute of Computing Technology, Chinese Academy of Sciences}
 \textsuperscript{3}\textit{Institute of Mathematics and Interdisciplinary Sciences, Xidian University}
 \textsuperscript{4}\textit{Huawei Cloud}
 \textsuperscript{5}\textit{Nanyang Technological University}
 }

\begin{abstract}
The Text-to-Video (T2V) model aims to generate dynamic and expressive videos from textual prompts. The generation pipeline typically involves multiple modules, such as language encoder, Diffusion Transformer (DiT), and Variational Autoencoders (VAE). Existing serving systems often rely on monolithic model deployment, while overlooking the distinct characteristics of each module, leading to inefficient GPU utilization. In addition, DiT exhibits varying performance gains across different resolutions and degrees of parallelism, and significant optimization potential remains unexplored.

To address these problems, we present \sysname, a flexible system that integrates both inter-phase and intra-phase optimizations. \sysname\ focuses on two key metrics: optimal degree of parallelism, which prevents excessive parallelism for specific resolutions, and starvation time, which quantifies the sacrifice of each request. To this end, \sysname\ introduces a \textit{decoupled control mechanism} to minimize the computational inefficiency caused by imbalances in the degree of parallelism between the DiT and VAE phases. It also designs a \textit{greedy resource allocation algorithm} with a novel scheduling mechanism that operates at the single-step granularity, enabling dynamic and timely resource scaling. Our evaluation on the T5 encoder, OpenSora SDDiT, and OpenSora VAE models across diverse datasets reveals that \sysname significantly outperforms state-of-the-art baselines by up to 1.44$\times$ in p99 latency and 1.43$\times$ in average latency.

\footnotetext{$^{*}$ Equal contribution.}

\end{abstract}

\maketitle

\section{Introduction} \label{sec: intro}
Since the advent of Sora from OpenAI \cite{url-sora}, text-to-video (T2V) generation becomes prevalent and progresses rapidly. Modern T2V solutions are normally based on the Diffusion Transformer (DiT) architecture, which outperforms conventional CNN-based methods \cite{blattmann2023stable} in generating high-quality images and videos. Despite their excellent performance, a significant challenge with DiT lies in their high computational demands, especially when generating high-resolution content. This is primarily due to the attention mechanism in the transformer, which has an $O(L^2)$ complexity relative to the input token length $L$, while DiT needs to process a large number of tokens in each denoising step \cite{yuan2024ditfastattn, chen2025pixart}.

It is important but challenging to develop an efficient T2V serving system with low latency and high resource utilization \cite{distrifusion-2024, li2024swiftdiffusion}. A number of techniques have been proposed to accelerate the video generation process, including distillation \cite{yuan2024ditfastattn}, post-training \cite{zhang2024cross}, and caching \cite{wimbauer2024cache}. Unfortunately, these methods exhibit some limitations, such as requiring additional training or compromising the output quality. A more promising solution, sequence parallelism \cite{liu2023ring, rethinking-icml20}, accelerates T2V serving by efficiently processing the rich spatial and temporal information. However, due to the distinct computational characteristics of different modules in T2V, as well as different serving requests, existing T2V systems suffer from large resource utilization inefficiency. We perform an in-depth investigation to disclose the following limitations.

{
\begin{figure}[t]
\begin{center}
\centerline{\includegraphics[width=0.48\textwidth]{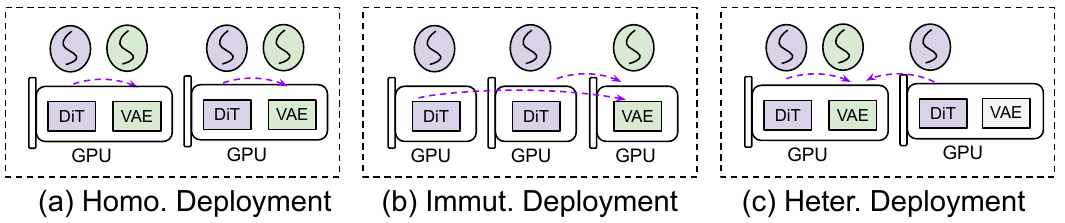}}
\captionsetup{belowskip=-0.8cm}

\caption{The deploy architecture of a T2V serving system.}
\label{fig-deploy-arch}
\end{center}
\end{figure}
}

First, \textbf{the monolithic model results in inefficient resource utilization, as it does not account for distinct computational requirements of the DiT and VAE components.} Specifically, as illustrated in Figure \ref{fig-deploy-arch} (a), existing systems employ a homogeneous sequence parallelism strategy across both DiT and VAE, with the same degree of parallelism (DoP). However, DiT and VAE exhibit heterogeneous sequence parallelism due to their different computational demands, even when processing the same request. For instance, as shown in Figure \ref{fig-dv-charac}, the computation time of VAE does not decrease per GPU as the DoP increases. Consequently, existing deployment fails to account for their distinct characteristics, leading to inefficient resource utilization. One potential solution, similar as the optimization of large language model (LLM) serving \cite{patel2023splitwise, hu2024inference, hu2024memserve, zhong2024distserve}, is to disaggregate the DiT and VAE phases and deploy them on independent GPU groups. As shown in Figure \ref{fig-deploy-arch} (b), each group deploys a DiT or VAE instance, tailoring the quantity to meet the specific requirements of the respective module. However, this immutable grouping strategy often struggles to adapt to the dynamic resource demands of different phases. The isolation among groups prevents elastic utilization of GPU computation, leading to large GPU wastage. 

Second, \textbf{existing T2V serving systems adopt the fixed deployment configuration to handle dynamical serving requests, leading to either huge resource wastage or prolonged latency.} 
DiT is the most time-consuming stage in T2V, involving multiple denoising steps. 
Its execution time depends on numerous factors including the resource allocation (e.g., number of GPUs), output format (e.g., video resolution, number of video frames), and model hyperparameters (e.g, number of denoising steps). Specifically, as shown in Figure \ref{fig-dv-charac}, to generate higher-resolution videos, a longer execution time is required, and it does not always decrease linearly as the DoP increases. Therefore, it is crucial to precisely configure the DoP for different resolutions. Unfortunately, existing solutions adopt the static DoP on a fixed amount of GPUs, failing to adapt to the distinct computational demands of different requests. A straightforward approach is to schedule requests based on their resolutions to optimize the GPU usage. While it can reduce the computational gap and improve the resource utilization, it fails to achieve precise alignment due to the fixed model configuration and diverse request loads. Moreover, existing scheduling algorithms operate at the request granularity, preventing adaptive resource reallocation during execution, as shown in Figure \ref{fig-fixed-utilization} (a). It remains an open question to identify the appropriate DoP for each specific resolution to achieve efficient scheduling.

To address the above limitations, we propose \sysname, an efficient T2V model serving system that decouples the model serving from the DoP of instance, enabling coordinated execution of DiT and VAE. It employs heterogeneous sequence parallelism and elastic resource allocation tailored to model characteristics, scheduling the execution of DiT at the granularity of \textit{steps} instead of \textit{requests}. The novelty of \sysname is reflected from the following two aspects. 

\ding{182}\,\textbf{Heterogeneous deployment.} We decouple DiT and VAE into independent modules and coordinate them heterogeneously as demands. Unlike existing homogeneous deployment or immutable grouping strategies, we propose a unified management mechanism that independently loads model weights and establishes connections. Thus, \sysname can effectively eliminate the resource isolation between immutable groups and dynamically adjust the number of GPUs allocated to each phase during execution. As shown in Figure \ref{fig-deploy-arch} (c), \sysname initially loads the model weights into each instance without pre-defined grouping. During the DiT phase, \sysname\ dynamically allocates two GPUs on demand to establish connections and execute the computationally intensive diffusion process. Upon completion of the DiT phase and transitioning to the VAE phase, \sysname\ reduces GPU usage to a single unit, effectively minimizing the resource redundancy.


\ding{183}\,\textbf{Stepwise on-demand scheduling.} We conduct performance profiling to identify the optimal DoP for each resolution, aiming to balance the resource utilization efficiency and latency. Accordingly, we schedule DiT execution at the granularity of a single step rather than an entire request, as shown in Figure \ref{fig-fixed-utilization} (b). This enables the serving system to process one model step at each time, resolving static allocation caused by engine initialization during processing. To this end, we design an engine controller to coordinate GPU resources and dynamically reschedule per step.

We perform extensive experiments to validate the effectiveness of \sysname. In particular, we compare \sysname with the state-of-the-art VideoSys \cite{videosys2024} when running the OpenSora model \cite{url-sora}. Our evaluation results show that by dynamically adjusting the inter-phase and intra-phase GPU resources, \sysname\ could reduce the p99 latency by 30.4\% and average latency by 30\% over heterogeneous deployment approaches.


The key contributions of this paper are as follows:
\begin{itemize}[noitemsep,topsep=0pt,leftmargin=*]
\item We conduct an in-depth empirical analysis of existing T2V serving systems in terms of resource utilization and performance characteristics, to disclose two key limitations.

\item To address the resource inefficiency between DiT and VAE caused by their distinct computational capacity, we develop an engine controller, enabling dynamic model weights loading and adaptive connections across various DoP configurations, optimizing the resource utilization.

\item To strike a balance between the optimal DoP of each request and reducing resource wastes caused by static deployment, we design a unified GPU mechanism to manage resources for model serving. We further propose a greedy algorithm for scheduling requests at the granularity of step instead of request, to maximize the GPU utilization. 

\end{itemize}

\section{Background}
\subsection{Diffusion Models}
Diffusion models have emerged as the dominant solution for image and video generation tasks, due to their exceptional ability to produce high-quality output \cite{peebles2023scalable}. In reverse diffusion, the process starts with pure Gaussian noise $x_T \sim \mathcal{N}(0, I)$, which is progressively denoised over $T$ iterative steps using the model $\epsilon_\theta$, ultimately generating the final result $x_0$. Specifically, given the noisy $x_t$ at each step $t$, $\epsilon_\theta$ predicts the noise $\epsilon_t$ based on the noisy input $x_t$, step $t$ and an additional input $t$ (e.g., text). The solver $\Phi$ calculates the output for the previous step using the following equation:
\begin{equation}
x_{t-1} = \text{$\Phi$}(x_t, t, \epsilon_t),   \epsilon_t = \epsilon_\theta(x_t, t, c)
\end{equation}
Diffusion models typically leverage a deep neural network, with U-Net \cite{ho2020denoising} or DiT as common backbones. Compared to U-Net, DiT is a more scalable transformer-based architecture, enabling greater model capacity \cite{pab-2024, xdit-2024}. The quality of the image or video, closely tied to its resolution, is critical for users \cite{url-KuaiShou, url-Ocean, url-runway}. However, higher resolutions lead to a quadratic increase in computational complexity, significantly affecting the generation latency \cite{yuan2024ditfastattn}.

\subsection{Text-to-Video Generation}
{
\begin{figure}[t]
\begin{center}
\centerline{\includegraphics[width=0.48\textwidth]{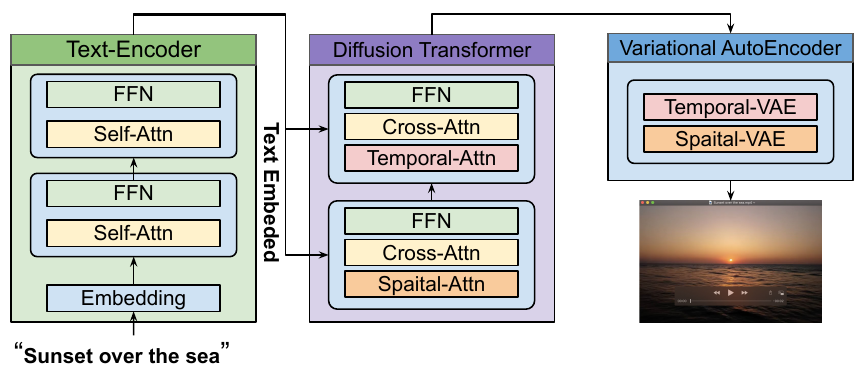}}
\captionsetup{belowskip=-0.8cm}

\caption{The architecture of T2V systems.}
\label{fig-tv2-arch}
\end{center}
\end{figure}
}

Text-to-Video (T2V) \cite{url-sora, zheng2024open, yang2024cogvideox} is a multimodal generation task that produces videos from text descriptions, often requiring integrating multiple collaborative modules. We use one of the most popular open-source T2V model, OpenSora \cite{lin2024open}, to illustrate the common architecture of this task. 

As shown in Figure \ref{fig-tv2-arch}, OpenSora comprises three critical modules. (1) The text encoder \cite{ni2021sentence, su2022one} transforms the input text into semantic feature representations, capturing the contextual and relational information to guide video generation effectively. (2) The DiT \cite{peebles2023scalable} generates the latent representations through a sequence of iterative denoising steps. The video denoising process reconstructs data from noise using three types of Transformer blocks: spatial, temporal, and cross-attention. Spatial blocks capture the relationships among tokens within the same temporal index; temporal blocks manage interactions across time; and cross-attention integrates the conditioning input to ensure coherence and alignment with the context. (3) The VAE \cite{pinheiro2021variational} decompresses latent representations into video frames, reducing the complexity while preserving the essential spatial and temporal features for efficient video generation. The primary operators in VAE include convolution, upsampling, and activation functions, with convolution contributing the most to the computational density. The convolution kernel involves generating high-resolution outputs, and its computational complexity is directly related to the resolution of the input feature map, the size of the convolution kernel, and the number of output channels \cite{li2021survey}.

\subsection{Model Serving Optimization}
Driven by the growing demand for AI applications, especially generative AI, model serving has become an important category of workloads in modern data centers, which are typically equipped with specialized accelerators such as GPUs or TPUs \cite{norrie2021design}. 
The response latency is a critical indicator for the performance of model serving. In T2V systems, the latency could be extremely long, especially for generating high-resolution videos. For instance, the generation time of one video by OpenAI Sora \cite{url-sora} can reach several minutes due to the time-intensive attention mechanism. Therefore, it is urgent to accelerate the model serving systems for better user experience and economic benefits. 


Numerous system-level optimization solutions have been designed targeting different types of AI tasks. Unfortunately, it is non-trivial to directly apply them to T2V systems. Specifically, batching is a critical technique in serving systems to achieve high resource utilization, particularly with GPUs, as it enables parallel processing of multiple inputs while minimizing the idle time \cite{yu2022orca, vllm-sosp23}. However, this approach is less effective in T2V models due to its high computational demands, as shown in Figure \ref{fig-dv-utilization}. Some studies leverage the similarities between consecutive steps to reduce the computational requirements, thereby optimizing the denoising process more efficiently \cite{selvaraju2024fora, ma2024deepcache, wimbauer2024cache}. However, cache reusing often compromises the quality of generated videos.

On the other way, many parallelism strategies have been designed to accelerate the inference of Transformer-based models, including Pipeline Parallelism (PP), Tensor Parallelism (TP), and Sequence Parallelism (SP) \cite{deepspeed-sc22, shoeybi2019megatron, li2021sequence}. TP is particularly effective for LLMs due to the large model sizes and relatively small activation sizes. The communication overhead introduced by TP is negligible compared to the latency reduction achieved through the increased memory bandwidth. LLMs utilize embedded SP by sharding along a single sequence dimension proposed in Ring-Attention \cite{liu2023ring}. In contrast, parallelism in diffusion models presents different challenges. These models are generally smaller than LLMs but often encounter bottlenecks due to large activation sizes, primarily caused by spatial dimensions, especially when generating high-resolution outputs. TP is unsuitable for DiT due to the substantial communication overhead during inference, while PP is ineffective because of the small model size. SP can benefit DiT in video generation, particularly in multi-dimensional Transformer-based models. However, the degree of SP must be carefully determined before deployment to achieve optimal performance and resource utilization.

\section{An Empirical Study}
\label{sec:empirical-study}
{
\begin{figure}[t]
\begin{center}
\centerline{\includegraphics[width=0.48\textwidth]{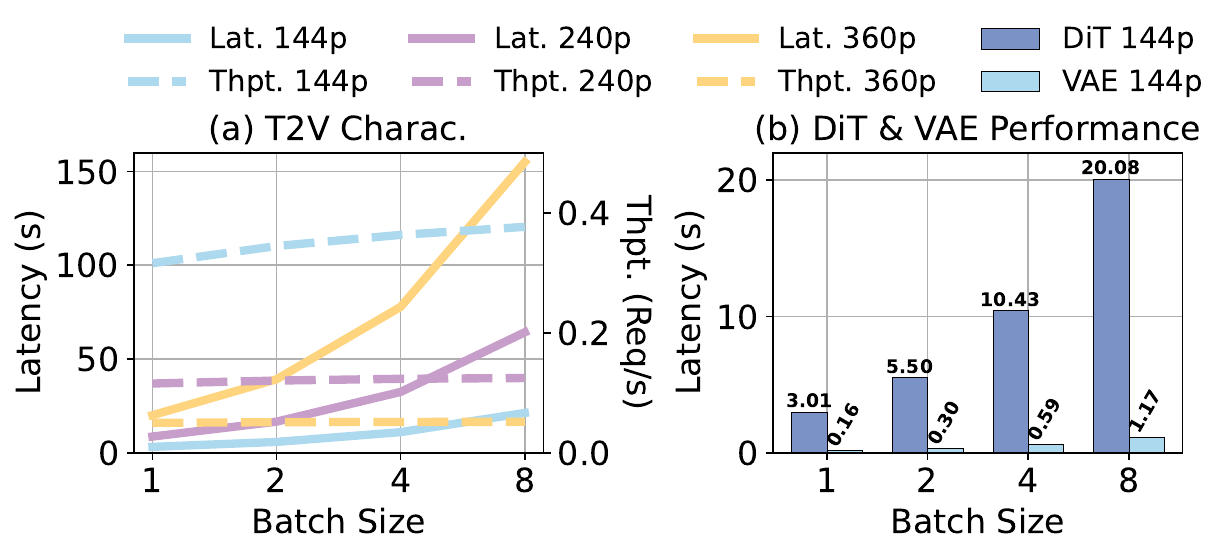}}
\captionsetup{belowskip=-0.8cm}

\caption{The impact of changing the batch size on the performance of T2V serving.}
\label{fig-t2v-charac}
\end{center}
\end{figure}
}
{
\begin{figure}[t]
\begin{center}
\centerline{\includegraphics[width=0.48\textwidth]{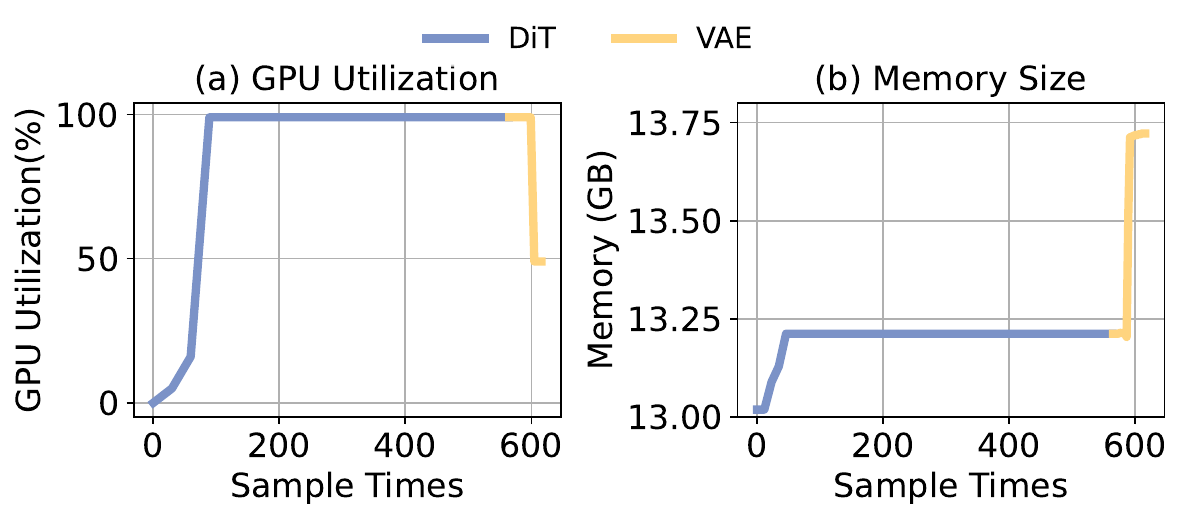}}
\captionsetup{belowskip=-0.8cm}

\caption{GPU Utilization of DiT and VAE (resolution: 144p).}
\label{fig-dv-utilization}
\end{center}
\end{figure}
}

We perform an in-depth investigation towards a representative T2V model, OpenSora \cite{lin2024open}. We discover some interesting findings, which can shed light on the design of new optimization approaches for T2V models. 

\subsection{Impact of Batch Size} \label{sec:batch size-impact}

We first measure the impact of the batch size on the serving inference. Figure \ref{fig-t2v-charac} shows the characterization results with the batch size ranging from 1 to 8 with the degree of sequence parallelism set to one. 
We observe the following pattern: in DiT, the throughput scales linearly with the batch size at smaller resolutions but plateaus as batch size increases. For larger resolutions, its remains nearly unchanged. 
In contrast, VAE maintains consistent system throughput across different resolutions and batch sizes.

We further evaluate the GPU utilization at runtime. As shown in Figure \ref{fig-dv-utilization}, both DiT and VAE saturate GPU resources with a batch size of 1, which is consistent with the behavior of attention and convolution operations. 
Batching can not improve the throughput of the T2V system but instead increasing the latency of requests within a single batch. 
Based on these observations, we have the following insight:

\begin{tcolorbox}[colback=gray!5!white,colframe=gray!75!black,left=1mm, right=1mm, top=0.5mm, bottom=0.5mm, arc=1mm]
    \textbf{Insight 1}: The optimal scheduling policy in T2V is to process requests sequentially, one at each time, because the GPU computational capacity is always limited.
\end{tcolorbox}



\subsection{Impact of DoP} \label{sec:sp-impact} 

Online video generation is a time-consuming process, and typically involves model deployment with multi-GPU parallelism. This approach distributes the computational workload to reduce the generation latency and improve the user experience. We evaluate the impact of the DoP (i.e., number of GPUs) on the T2V system. We set the DoP from 1 to 8 in our generation tests. 

As shown in Figure \ref{fig-dv-charac}, requests with larger resolutions demonstrate distinct characteristics across two phases under varying degrees of parallelism: in DiT, higher-resolution requests distribute the computational load across more GPUs as the DoP increases, resulting in an initial linear reduction in execution time, followed by a slower rate of decrease. However, VAE maintains stable execution time regardless of GPU parallelization. As the DoP increases, the latency of VAE accounts for a larger proportion of the total serving time, leading to greater resource inefficiency with more GPUs. For example, as shown in Figure \ref{fig-dv-charac}, 
the VAE execution time accounts for 14.3\% of the total inference time with four GPUs, compared to 4.5\% with a single GPU under the request with 360p resolution. We find that the key issue lies in all GPUs within the VAE processing the same input tensor from DiT, despite utilizing the same number of GPUs as DiT during generation. 
Therefore, with DoP=4, VAE wastes the computational time of three GPUs. This redundancy highlights inefficiencies in the current homogeneous-deployment manner. Furthermore, VAE cannot achieve acceleration, even when using parallelized versions such as DistVAE \cite{xdit-2024}.

According to the above discussion, the monolithic T2V serving performs DiT and VAE sequentially, ignoring the inconsistent hardware resource allocation between phases, resulting in redundant computation during inference. This gives us the following design insight: 

\begin{tcolorbox}[colback=gray!5!white,colframe=gray!75!black,left=1mm, right=1mm, top=0.5mm, bottom=0.5mm, arc=1mm]
    \textbf{Insight 2}: We need to decouple DiT and VAE from model deployment and dynamically allocate resources to each phase for improved efficiency.
\end{tcolorbox}

In Appendix, Figure \ref{fig-total-charac} further demonstrates the effects of batch size and DoP on the T2V models are orthogonal.

{
\begin{figure}[t]
\begin{center}
\centerline{\includegraphics[width=0.48\textwidth]{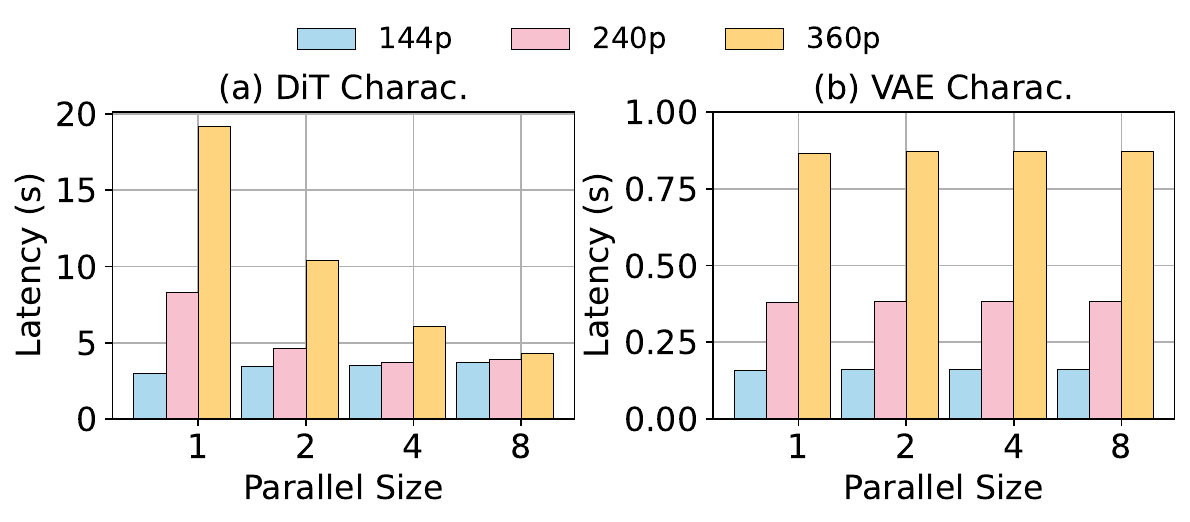}}
\captionsetup{belowskip=-0.8cm}

\caption{Latency of DiT and VAE under various DoPs.}
\label{fig-dv-charac}
\end{center}
\end{figure}
}

\subsection{Impact of Video Resolution} \label{sec:motivation-resolution}

Most T2V serving systems allow users to define the resolution through a configurable parameter, as seen in Kling \cite{url-KuaiShou}, Video Ocean \cite{url-Ocean}, and Runway \cite{url-runway}. Therefore, T2V serving systems must handle workloads consisting of requests with varying resolutions. The latency of a request increases with higher resolutions, and DiT acts as a significant contributing factor. To handle the specific resolution efficiently, a optimal number of GPUs should be allocated to ensure acceptable latency for users and maximize resource utilization. 

We evaluate the impact of varying resolutions, ranging from 144p to 360p, on the DiT performance across parallel sizes from 1 to 8. The results are shown in Figure \ref{fig-dv-charac}. During the DiT phase, we observe two key patterns: (1) for lower-resolution requests, the performance remains stable across varying DoP configurations, with slight degradation at higher DoP; (2) for higher-resolution requests, the performance improves significantly as DoP increases. For instance, a request with 144p resolution performs better on a single GPU than on multiple GPUs. Similarly, deploying a 240p resolution request on two GPUs can halve the latency compared to one, but it cannot get enough earnings for more than two GPUs because of Amdahl's law. 
As shown in Figure \ref{fig-opt-gpu-num}, when doubling the DoP, the reduction ratio in DiT execution time is different under different initial DoP values. 
It highlights that performance gains do not scale linearly with the GPU count. 
This inspires us as follows:


\begin{tcolorbox}[colback=gray!5!white,colframe=gray!75!black,left=1mm, right=1mm, top=0.5mm, bottom=0.5mm, arc=1mm]
    \textbf{Insight 3}: Larger DoP does not always lead to better performance. It is necessary to determine an optimal configuration, enabling efficient request scheduling to maximize the performance gain for each resolution.
\end{tcolorbox}






\subsection{Impact of Static Deployment}

In existing T2V serving systems, model deployment remains static on a fixed amount of resources throughout the entire process, and each request runs continuously until completion. This is inefficient and leads to significant computational waste due to the mismatch between resource allocation and optimal requirements. For example, Figure \ref{fig-fixed-utilization} illustrates the idle resource time in a static scenario with four GPUs serving two requests with resolutions of 240p and 360p. When a 240p request is followed by a 360p request, the scheduler allocates two GPUs to the first request, leaving another two idle. This does not satisfy the optimal DoP requirement for the subsequent request. The scheduler then faces two choices: (1) waiting for the 240p request to finish and release its GPU, leading to wasted idle resources, (2) allocating the two idle GPUs to the 360p request, causing it to execute suboptimally with insufficient resources until completion, as it cannot utilize the additional two GPUs released by the 240p request due to the static allocation at initialization.



{
\begin{figure}[t]
\begin{center}
\centerline{\includegraphics[width=0.48\textwidth]{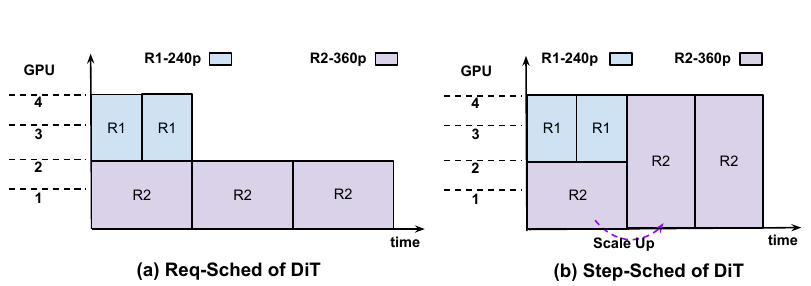}}
\captionsetup{belowskip=-0.8cm}

\caption{The idle GPU under the fixed deployment.}
\label{fig-fixed-utilization}
\end{center}
\end{figure}
}

According to the above analysis, static model deployment and coarse-grained request scheduling lead to inefficient resource utilization and substantial wastage. This observation provides the following design motivation: 

\begin{tcolorbox}[colback=gray!5!white,colframe=gray!75!black,left=1mm, right=1mm, top=0.5mm, bottom=0.5mm, arc=1mm]
    \textbf{Insight 4}: It is necessary to design a fine-grained, step-aware scheduling mechanism and flexible model management for dynamic resource scaling. These enable elastic parallelism deployment for DiT, dynamically adjusting based on available resources during request execution.
\end{tcolorbox}

\section{\sysname Design}
{
\begin{figure}[t]
\begin{center}
\centerline{\includegraphics[width=0.48\textwidth]{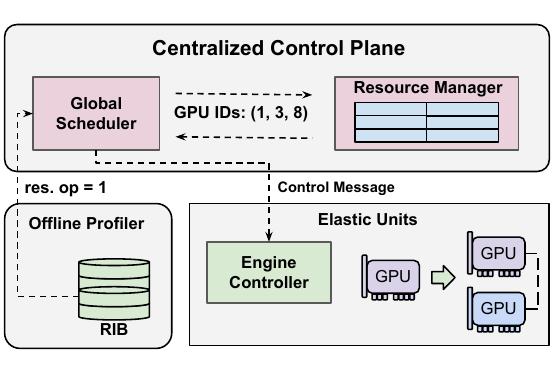}}
\captionsetup{belowskip=-0.8cm}

\caption{Architecture overview of \sysname.}
\label{fig-ddit-arch}
\end{center}
\end{figure}
}

Inspired by the insights from our empirical study, we design \sysname, a novel T2V serving system that can efficiently allocate GPU resources at each phase, and flexibly scale the resources for running requests at a fine-grained step level. 

\subsection{Overview}
The design of \sysname encompasses the following ideas. First, we decouple the DiT and VAE phases for deployment and execution independently. Second, we decouple the model weights loading and the establishment of communication groups for parallelism in itialization. The model weights are loaded into each GPU instance, and communication groups are established on demand, eliminating the need to predefine the number of GPUs. Third, we determine the optimal DoP for various resolutions in the offline manner, which are used to guide the online resource allocation. Fourth, we implement a greedy resource allocation algorithm with a step-wise scheduling strategy to minimize the idle GPU time and efficiently change DoP for handling requests. 
Figure \ref{fig-ddit-arch} shows the architectural overview of \sysname, consisting of three main components: Offline Profiler, Centralized Control Plane, and Engine Units.

\noindent\textbf{Offline Profiler.} This module automatically analyzes the characteristics across various request resolutions and applies the predefined policies to determine the optimal DoP. It is executed only once for each unique resolution. The pair of resolution and optimal DoP value is stored in a request information database (RIB). The resolution must be profiled first if its portrayal is not available.

\noindent\textbf{Centralized Control Plane.} It consists of a global scheduler, a cluster monitor, and a resource allocator. The global scheduler distributes requests to DiT instances and retrieves outputs from VAE instances. The cluster monitor tracks the GPU status and collects request statistics from Engine Units. The resource allocator manages GPU allocation, treating each GPU as a fundamental unit that contains a full replica of the model weights, enabling a heterogeneous model parallelism strategy. The global manager dynamically determines the DoP for requests based on available resources and their optimal DoP before each step of the DiT and VAE phases.


\noindent\textbf{Engine Units.} The Engine Units consist of an engine controller and a model engine. The engine controller receives control messages from the global scheduler, manages connections between instances using GPU IDs, and triggers the model engine to execute step by step. It adjusts the communication groups based on updated DoP, broadcasting latent representation tensors to newly assigned instances. It also periodically reports the request phase, number of steps, and resource usage to the cluster monitor to aid GPU allocation.

Below we elaborate two core approaches in \sysname.
\subsection{Global Scheduler}
\sysname\ features a scheduler designed for efficient request dispatching to the DiT phase in an online serving system. Traditional scheduling focuses only on the instance load, which is unsuitable for T2V serving as it overlooks the unique characteristics of T2V serving. We propose a new scheduling algorithm that accounts for the computational requirements of requests in DiT, enabling efficient handling of varying resolutions. To maximize GPU utilization, we design a step-wise mechanism for the DiT engine to avoid waiting time.

\subsubsection{Problem Formulation}

The objective of the global scheduler is to minimize the overall processing time of pending requests given their volume and available resource capacity, which is also equivalent to minimizing the cumulative resource occupancy time, defined as:
\begin{equation}
  O=\sum_{GPU_{j}\in cluster} occupied\_time(GPU_{j})  
\end{equation}
In addition to the scheduling method, the characteristics of requests, such as resolution distribution, also play a key role in influencing this metric, as analyzed in Section \ref{sec:motivation-resolution}. We first define the resolution distribution of a request batch comprising $N$ resolution types, with each type having a corresponding proportion $x_i$, where $1\le i\le N$. Then we present a scheduling algorithm below that derives the theoretical minimum of this metric under the fixed resolution distribution and request volume constraints.

\subsubsection{Theoretical Optimal Scheduling Algorithm} \label{sec:theroetical-optimal-algorithm}

The optimal scheduling algorithm can theoretically be achieved if the resolution distribution is already known. Under this assumption, we use dynamic programming to determine the optimal resource allocation for requests of each resolution type. We present the details of scheduling in Algorithm \ref{Theoretical Upper Bound Solver}.

\begin{algorithm}
\footnotesize
\caption{Theoretical Optimal Scheduling Algorithm}
\begin{algorithmic}[1]
\State $m,\ n\gets$ numbers of instances and GPUs per instance, $M = m\times n$
\State $N \gets $ numbers of resolution types
\State $dp[M+1][N+1] \gets [[0,0,...], [0,0,...],...]$ 
\State $dp[i][0]_{1\le i\le M} =0,\ dp[0][j]_{1\le j\le N} = \infty$
\State $x_{1}:x_{2}:x_{3}...x_{N} \gets $ the proportion of each resolution type
\State $ps=\{1,2,4,...\} \gets Process\_Group\_Size\_List$
\State $G[m][n] \gets Allocated\_GPU\_List$

\For{$i = 1$ to $M$}
    \For{$j=1$ to $N$}
        \State \text{FIND\_OPTIMAL\_TIME}($G$, $dp$, $ps$, $i$, $j$, $x_{j}$)
    \EndFor
\EndFor
\State \textbf{Return} $dp[M][N]$ 
\Function{FIND\_OPTIMAL\_TIME}{$G$, $dp$, $ps$, $i$, $j$, $x_{j}$}
   \For{$k=1$ to $i$}
        \For{$p$ in $ps$}
            \State $\alpha \gets \textbf{BandwidthAwarePartition}(\underset{(i-k+1)_{th}\sim i_{th}}{G(net.)},k,p)$
            \If{$ \alpha == 0$}
                \State \textbf{Continue}
            \EndIf
            \State $d \gets \textbf{EstimateExecutionTime}(p, j)$
            \State $dp[i][j]={\min}(dp[i-k][j-1]+k\times \textbf{Occupy}(x_{j}, d, \alpha))$ 
        \EndFor
    \EndFor
\EndFunction

\end{algorithmic}
\label{Theoretical Upper Bound Solver}
\end{algorithm}

\par The objective is to find the minimum cumulative resource occupancy time of assigning total $M$ GPUs to total $N$ resolution types of requests. 
We assume that there are $m$ instances and each instance contains $n$ GPUs (line 1). 
We adopt dynamic programming, where $dp[i][j]$ is used to represent the optimal result of assigning the first $i$ GPUs to the first $j$ request types (line 3) and initializing it (line 4).
Since the request execution time varies across different DoP values, we also enumerate all the values (line 6) and use $G[i][j]$ to represent whether the $j$-th GPU in the $i$-th instance is allocated or not (line 7). 

\par Subsequently, the algorithm calculates each $dp[i][j]$ using a double loop (line 8 $\sim$ 10). 
The function starts at line 12 is the core logic for calculating $dp[i][j]$.
It enumerates the number of GPUs and the DoP used to execute the $j$-th request type (line 13 $\sim$ 14). 
Line 15 calculates the number of model instances ($\alpha$) based on network bandwidth conditions between the $(i-k+1)$-th $\sim$ $i$-{th} GPUs and DoP ($p$). 
Since SP requires high-speed inter-GPU connectivity, the allocation strategy varies depending on the cluster configuration. 
For example, considering a cluster of two machines, each is equipped with eight GPUs, where NVLink enables high-speed bandwidth within machines but only low-speed bandwidth exists between machines.
If the first machine and the first GPU of the second machine are already occupied, allocating an additional seven GPUs yields different outcomes based on the DoP. With the DoP of one, seven model instances can be created if each GPU hosts the full model weights. However, if the DoP is four, only one model instance can be created. 
\par Since the execution time of the T2V model is deterministic once parameters such as denoising steps, DoP ($p$) and resolution ($j$) are fixed, we pre-profile across various scenarios and leverage these data to estimate the execution time ($d$) for a single request with the given resolution type $j$ in line 18. Line 19 calculates $dp[i][j]$ by the minimal sum of the two parts: (1) the cumulative resource occupancy time for assigning the first $(i-k)$ GPUs to the first $(j-1)$ request types, and (2) the cumulative resource occupancy time for assigning the $(i-k+1)$-th $\sim$ $i$-{th} GPUs to the $j_{th}$ request type. 
\par The first part is already calculated by dynamic programming, and the second part can be modeled as a batch model or a queue model, depending on the assumptions:
\begin{itemize}[noitemsep,topsep=0pt,leftmargin=*]
    \item \textbf{Batch Model}:
    If there are already $S$ requests pending in the system before it starts processing and no more new requests will arrive. 
    Given that the number of model instances $\alpha \ge 1$ and the execution time $d$ is fixed, we can model the scenario using a batch model. 
    Assuming that the request is evenly distributed across $\alpha$ model instances, the number of type $j$ requests executed by each instance is $\lceil \frac{S \cdot x_{j}}{\alpha} \rceil$. 
    The execution time of request type $j$ is $d$, so the average resource occupancy time for a single GPU is computed as follows:
    \begin{equation} \label{batch-model}
    W_{Batch}(type_{j})= \lceil \frac{S \cdot x_{j}}{\alpha} \rceil d
    \end{equation}
    Therefore, the cumulative resource occupancy time for assigning the $(i-k + 1)$-th $\sim$ $i$-th GPUs to the $j_{th}$ request type is obtained by $k$ $\times$
    e.q. \ref{batch-model}.
    \item \textbf{Queue Model}: 
    This model characterizes a dynamic equilibrium scenario where the system maintains a constant pending requests count $S$, balancing continuous request arrivals with concurrent processing throughput. Due to space constraints, the detailed specification of the queue model is provided in Appendix \ref{appendix-a}. 
\end{itemize}

In practical scenarios, the resolution distribution is not known in advance. Previous works\cite{yang2024preact, poppe2020seagull} attempt to predict the request loads and their characteristics based on historical data patterns. 
However, these works exhibit limited robustness to high load fluctuations, and the runtime overhead induced by frequent resource re-allocations in response to load variations remains a critical under-addressed challenge. 
Therefore, to reduce the cumulative resource occupancy time in the real world, we introduce the greedy scheduling algorithm of \sysname.

\subsubsection{Greedy Scheduling in \sysname} \label{sec:b value}

\sysname schedules each request sequentially in a First-Come-First-Serve (FCFS) manner, since the batching technique is not suitable for T2V models as shown in Section \ref{sec:batch size-impact}. 
More specific, it employs a greedy approach to set the DoP of each request before the DiT phase and automatically choose whether to promote the DoP at a step granularity during the DiT phase, in order to minimize the cumulative resource occupancy time as well.

\par According to Amdahl's Law \cite{amdahl}, the optimal DoP for any resolution constant at one if we aim to minimize the cumulative resource occupancy time at a single request level. 
However, this configuration of parallelism severely degrades the performance for high-resolution requests and blocks subsequent ones. 
Therefore, we need to reassess the optimal DoP for each resolution, balancing the resource efficiency and performance.

\par As the DoP of the request increases, the proportional of time reduction in DiT phase afforded by each doubling of DoP is variable. 
In Figure \ref{fig-opt-gpu-num}, we present the change rate of the per-step DiT execution time between two adjacent DoP across various resolutions. We define this metric below:

\begin{equation} 
z = 1-\frac{DiT_{step\_time}(parallel\_size=i,\ r)}{DiT_{step\_time}(parallel\_size=i/2,\ r)}
\end{equation}
The $r$ means the resolution (144p, 240p, 360p) of the request and $i=2,4,8$ in Figure \ref{fig-opt-gpu-num}. 
For each request type we denote the value of $i$ as $B$ when the corresponding $z$ is largest. 
Therefore, for requests with resolutions of 144p, 240p, and 360p, their corresponding $B$ values are 1, 2, and 4, respectively.
Notably, we only treat $B$ of each request type as its optimal DoP for the DiT phase, while the optimal DoP is much smaller during the VAE phase, as analyzed in Section \ref{sec:sp-impact}. Next, we introduce how \sysname specifically applies the greedy strategy and the $B$ values to schedule requests, 

\par The first greedy scheduling occurs when a request arrives in \sysname and attempts to acquire $B$ GPUs, but only $G$ free GPUs are available ($G<B$). In this case, \sysname greedily assigns these $G$ GPUs to the request, allowing it to start DiT execution. The second greedy scheduling happens during the DiT phase if newly released GPUs become available from completed requests. If so, \sysname greedily assigns these GPUs to the request, allowing it to reach $B$ GPUs. This reassignment occurs only if the request has the highest priority among those requests whose current DoP is lower than its corresponding $B$. The definition of the priority lies below:
\begin{equation}
r_{starv.}+=(r_{cur\_step}-r_{last\_step})\times(r_{cur\_step\_time}-r_{opt\_step\_time})
\label{eq-hug-priority}
\end{equation}
It represents the theoretical additional DiT execution time that a request experiences from its most recent GPU assignment event until the current scheduling event, due to an insufficient number of GPUs (i.e., fewer than $B$). We refer to this measure as the starvation time, and the longer the starvation time, the higher the scheduling priority. More details are presented in Appendix \ref{sec:appendix-b} due to space constraint.

{
\begin{figure}[t]
\begin{center}
\centerline{\includegraphics[width=0.48\textwidth]{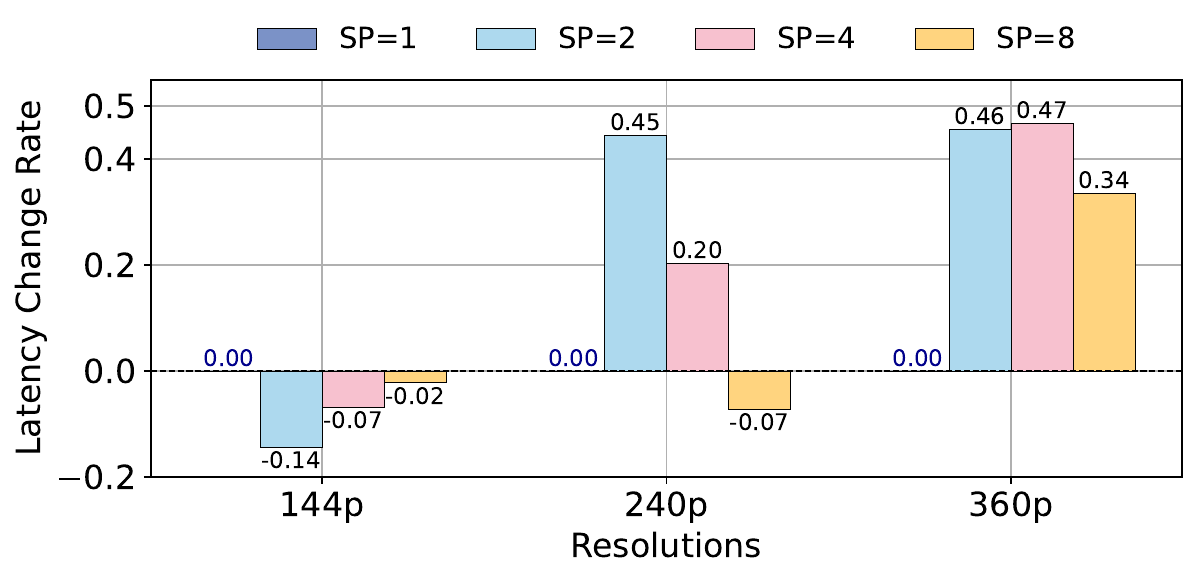}}
\captionsetup{belowskip=-0.8cm}

\caption{Per-step DiT time change rate between adjacent parallel sizes.}
\label{fig-opt-gpu-num}
\end{center}
\end{figure}
}

\subsection{Efficient Decoupling} 

\sysname\ introduce two decouple mechanisms: inter-phase and intra-phase designs in the T2V system. The inter-phase decoupling is necessary because the coupling module overlooks the different resource requirements of components like the DiT and VAE modules. \sysname\ decouples DiT and VAE into two independent components, enabling them to execute sequentially within the same instance at any time, as long as the VAE phase can utilize the latent representations tensor generated by the DiT phase. 

To optimize resource usage, \sysname\ employs an engine controller to scale down the DoP of the Engine Unit after completing the DiT phase. Specifically, for an Engine Unit $R$ with a DoP $d$, the elastic scale-down mechanism adjusts it to a new parallel group $\acute{R}$ with a reduced DoP $\acute{d}$, determined by the resource requirements of the VAE model, where $\acute{d}<d$ in \sysname. The mechanism designates the GPUs with the lowest IDs as master units within Engine Unit $R$, which remain in $\acute{R}$ during the VAE phase. The latent representations tensors are retained in these master units after the DiT phase completion. 

Moreover, the scale-down mechanism involves two overlapping Engine Units, each with different GPUs. They should use separate communication handlers to ensure message consistency. However, the coupling model necessitates reloading the model weights and rebuilding the communication domain. To address this, \sysname\ separates model weights loading and parallelism communication construction in model serving. Model weights are loaded initially on all GPUs, while parallelism communication domain construction is deferred until the beginning of the DiT or VAE phase.

\sysname\ proposes the intra-phase decoupling because the DoP configuration for the DiT phase is predefined and cannot change across multiple denoising steps. \sysname\ employs an engine controller to track each step of the engine units through inter-process messaging. It is aware of the DoP configuration at each step and receives updated DoP messages from the global scheduler, enabling flexible DoP reconfiguration, as discussed in \S\ref{sec:b value}.

{
\begin{figure}[t]
\begin{center}
\centerline{\includegraphics[width=0.48\textwidth]{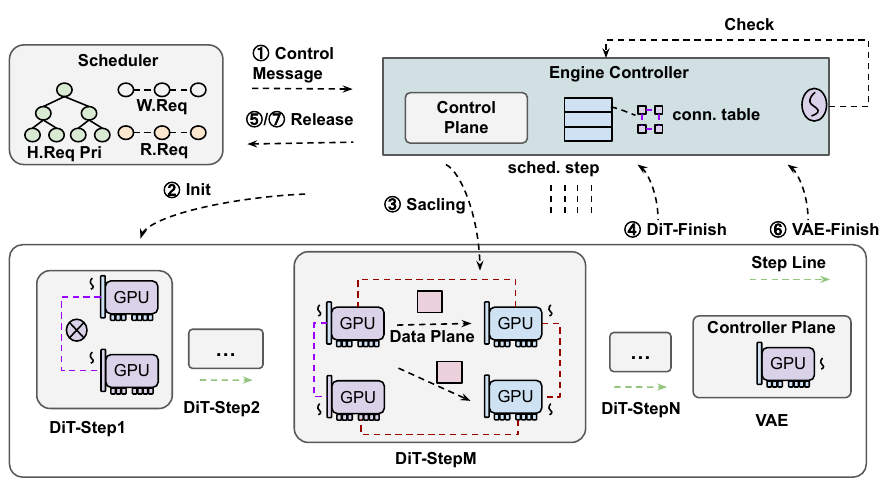}}
\captionsetup{belowskip=-0.8cm}

\caption{The lifecycle of a request in \sysname.}
\label{fig-lifecycle-req}
\end{center}
\end{figure}
}

Figure \ref{fig-lifecycle-req} illustrates the execution lifecycle of a request in the \sysname, emphasizing resource scaling and step scheduling. We still assume that when the request $b$ arrives, the available resource $G$ in \sysname does not meet its optimal requirement $B_{b:res.}$, as the request $a$ is still running. The scheduler retrieves $b$ from the waiting queue and attempts to allocate $B_{b:res.}$ GPUs through the resource manager, but only $G$ is allocated. Request $b$ is labeled as "hungry" and forwarded to the engine controller with a control message containing GPU IDs and request metadata. 

The engine controller searches the torch distribute handler using the hash value of the GPU IDs in the connection table. If no handler exists, it first triggers the workers (a worker represents a process executing T2V generation on a GPU) associated with the $G$ GPUs to establish the connection for parallelism and execute the DiT model by using the request metadata.
When $a$ completes, its resources $G_{a}$ are released. The scheduler can reassign these resources to $b$, and send an updated message with the new GPU IDs to the engine controller. The engine controller employs a background thread to monitor notifications. Upon detecting the available resources, it promotes the parallelism group to include the newly assigned GPUs and uses NCCL to transmit intermediate tensors to the new GPU workers. This scaling mechanism continues until DiT completes or the optimal number of GPUs is reached.

The engine controller communicates with workers at each step. Once the DiT phase finishes, it promptly receives the completion information. It stores the handler, releases partial GPUs, and notifies the scheduler. It also rebuilds the parallelism group for the VAE module. The remaining GPUs continue to execute until the the request is completed.


\textbf{Discussion:} 
We now discuss two auxiliary components in the T2V system: the text encoder and the tensor-to-video converter. (1) The text encoder currently requires negligible processing time, but it can be deployed on dedicated hardware for longer texts or multiple inputs. We leave this optimization for future work. (2) The tensor-to-video converter runs synchronously on the CPU in VideoSys \cite{videosys2024}, preventing GPU task blockage on the same machine. However, since it can be designed as an asynchronous process, we exclude this component from the experiment to ensure that all optimizations focus on GPU-executed tasks.
\section{Implementation}
We implement the Offline Profiler and \sysname’s Centralized Control Plane from scratch in Python. We adopt DiT and VAE instances based on Videosys \cite{pab-2024} and provide a FastAPI front end of \sysname\ for user convenience, allowing users to submit requests with customizable parameters such as resolution and aspect ratio. Each GPU runs a dedicated Python process to handle a portion of the sequence parallelism. We use RPC \cite{srinivasan1995rpc} for the communication in the control plane of \sysname, while NCCL \cite{url-nccl} is used in the data plane between Engine Units.

The Offline Profiler automatically conducts tests for specific resolutions with the increasing parallelism size to determine the optimal GPU count, storing the profiled data in MySQL \cite{dubois2013mysql}. The Centralized Control Plane comprises a resource allocator and a global scheduler. The resource allocator employs multi-level lists to organize GPUs into a buddy system, which manages GPU pairs for various DoP by automatically merging and splitting them as needed. The global scheduler retrieves profiling data from MySQL \cite{dubois2013mysql} during initialization, records the status of GPUs using a bitmap, manages communicator handlers in a hash table, and monitors the execution phase of requests. It uses RPC \cite{srinivasan1995rpc} to communicate with the elastic units to notify the scheduling information. The master unit employs NCCL \cite{url-nccl} to broadcast the DiT tensors to news, dynamically scaling sequence parallelism as needed. A DiT or VAE instance is a fundamental deployable unit, which has three coroutines in a Python process that runs the local scheduler, the reporter, and the main engine.



\section{Evaluation}
We evaluate \sysname using multiple real-world solutions across different workloads, quantify the effectiveness of its components, and further analyze its scalability against the theoretical optimal scheduling algorithm.

\subsection{Setup} \label{sec-setup}
\begin{table}
    \centering
    \begin{tabular}{ | p{1in} | p{0.8in} | c |  }
    \hline
       Name & Model & Param  \\
    \hline
       Text Encoder &  T5v1.1-xxl & 4.8B  \\
    \hline
       Diffusion Model &  STDiT3 & 1.1B \\
    \hline
       Var. Encoder &  OpenSoraVAE & 384M \\
    \hline
    \end{tabular}
    \caption{Detailed module parameters in our T2V system.}
    \label{table-configuration}
\end{table}

\noindent\textbf{Model.} We adopt the T5v1.1-xxl Encoder \cite{t5-v1_1-xxl} as the text encoder, STDiT-v3 \cite{STDiT-v3} as the DiT model, and OpenSoraVAE \cite{OpenSora-VAE} as the VAE model in our evaluation. These models are widely used in practice, with their parameters are detailed in Table \ref{table-configuration}. 

\begin{figure*}[t]
\centering
\centerline{\includegraphics[width=\textwidth]{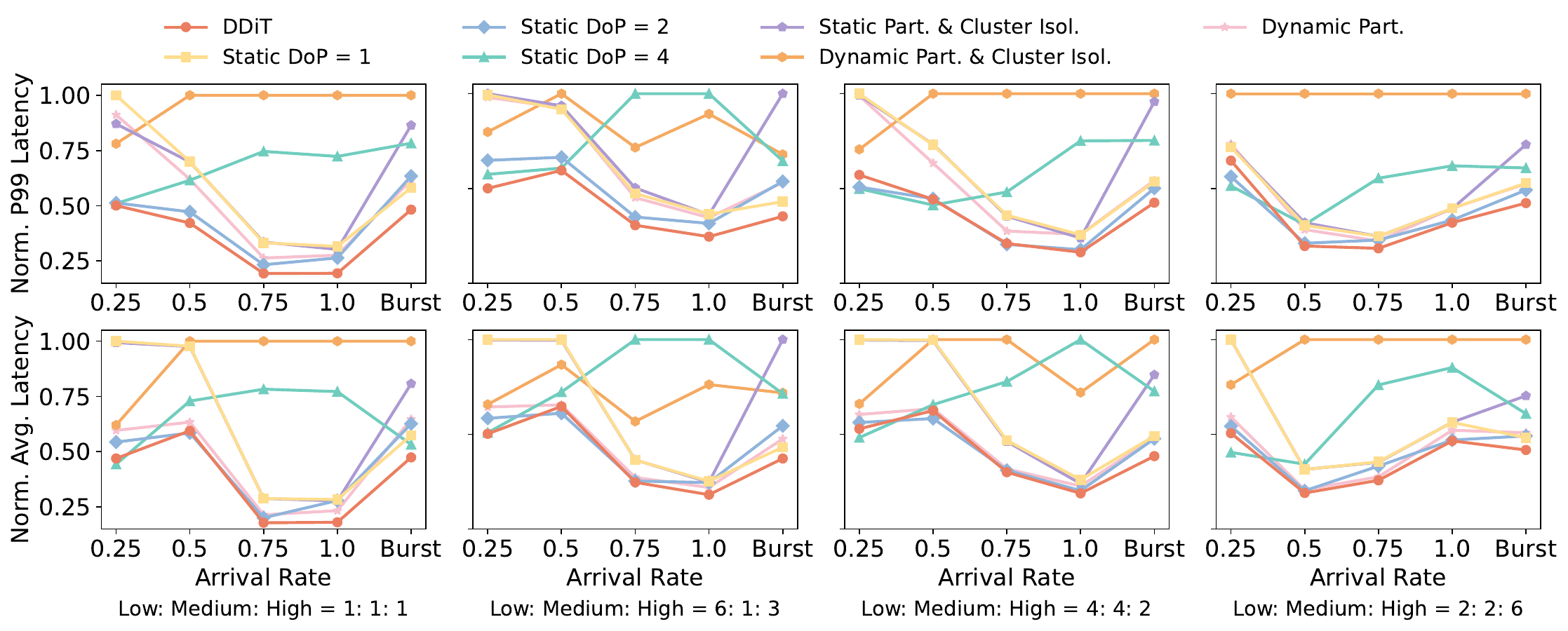}}

\caption{Single-node end-to-end performance (Lower latency indicates better performance). Part is an abbreviation for Partition, and Isol is isolation for short.}
\label{fig-total-comparison}
\end{figure*}

\noindent\textbf{Testbed.} We evaluate \sysname on a single server equipped with an Intel Xeon Gold 5218R CPU (192 cores and 2 TB of memory) and eight NVIDIA H800 GPUs (each with 80 GB of HBM), interconnected via a 400 GB/s NVLink \cite{url-nvlink}. The server runs Ubuntu 20.04, CUDA 12.1, Python 3.8, and PyTorch 2.2.1. For scalability experiments, we emulate \sysname and the baselines on a GPU cluster of eight servers, interconnected via 200 Gbps RDMA links.

\noindent \textbf{Workloads.} We evaluate the effectiveness of the greedy scheduling algorithm and resource allocation strategies in \sysname for T2V inference by varying the resolution while keeping the number of frames constant at 51 and denoising steps set to 30 as discussed in Section \ref{sec: intro}. Since no open-source real-world T2V workload trace was available, we selected 144p, 240p and 360p to represent low, medium and high resolutions, respectively, and adjusted their proportions to emulate a realistic workload. The requests arrive randomly as $A(t)\sim \lambda e^{-\lambda}$, where the arrival rate ($\lambda$) ranges from 0.25 and 1 to simulate varying load intensities. The burst in Figure \ref{fig-total-comparison} represents requests arriving simultaneously, simulating an extreme load.

\noindent\textbf{Baselines.} We use VideoSys \cite{videosys2024} as our baseline, the most popular open-source T2V system. We enhance it to support the following four real-world scenarios and the theoretical optimum scheduling algorithm in \S\ref{sec:theroetical-optimal-algorithm}.


\begin{itemize} [noitemsep,topsep=0pt,leftmargin=*]
\item \textbf{Static DoP(SDoP):} 
The DoP for the model instance is pre-configured and remains fixed to serve requests with varying patterns in practical environments, ensuring ease of deployment. We set the DoP range from one to four.

%
\item \textbf{Static Partition \& Cluster Isolation(SPCI):} 
T2V inference services involve multiple types of requests, with their ratios varying dynamically. To effectively serve specific request types, GPU resources are typically partitioned into clusters based on the historical patterns of request types, each with a fixed DoP. In the experiment, we allocate GPUs into three clusters based on predefined ratios of request types to optimize performance, assuming prior knowledge of the request distribution.

%
\item \textbf{Dynamic Partition \& Cluster Isolation(DPCI):} 
The dynamic partition prioritizes service quality compared to the static partition, as each request type has different computational demands. Specifically, we configure the DoP of clusters corresponding to each request type, as introduced in \ref{sec:b value}. In the experiment, we allocate an equal number of engine units for the three request types in their respective clusters, limited by the available hardware.

%
\item \textbf{Dynamic Partition(DP):} 
It removes the strict routing constraint to enhance the flexibility of dynamic partitioning in handling workload distribution shifts. For example, a $B=4$ request facing resource saturation in its native cluster can be adaptively downgraded to a $B=2$ cluster, minimizing resource idle time and ensuring service continuity.
\end{itemize}

The $B$ values we used in \sysname are introduced in Section \ref{sec:b value} and the DoP for the VAE phase is set to one according to Figure \ref{fig-dv-charac}.

\par \textbf{Metrics.} 
We focus on the average latency, the p99 latency, and the monetary cost of a T2V inference system. For each arrival rate, we compare different systems by normalizing their metrics. For example, at a given arrival rate, the normalized p99 latency for each system is calculated through dividing its p99 latency by the maximum p99 latency observed among all systems. The same normalization applies to the average latency and the monetary cost as well. Moreover, we treat the monetray cost numerically the same as the cumulative resource occupancy time in the T2V system, as described in Algorithm \ref{Theoretical Upper Bound Solver}, under the assumption of a constant charge of one unit per second per GPU. 



\subsection{End-to-End Performance}
\par \textbf{Single-Node Performance.} We compare \sysname\ with four baselines across ten pattern workloads using two key metrics. DDiT outperforms all other systems in every scenario. Results for five groups are shown in Figure \ref{fig-total-comparison}, while the remaining results are provided in Appendix \ref{sec:appendix-e2e} due to space limitations.

When the arrival rate is low (e.g., 0.25 or 0.5), the lifetimes of two consecutive requests do not overlap in underutilized systems, and allocating more GPUs enhances performance. As a result, the static DoP approach achieves the better performance across all four baselines with different DoP values.

\par At an arrival rate of 0.25, the optimal static DoP value is 4, while \sysname further reduces p99 latency by 12.88\%. When the arrival rate increases to 0.5, the optimal DoP value decreases to 2. \sysname can also achieve a 10.71\% reduction in p99 latency. This improvement is attributed to the DiT-VAE decouple mechanism, which accelerates resource release and enables subsequent requests to start execution earlier. The average latency increases by 9.43\% or remains similar to the optimal static DoP at each arrival rate, as the limited number of GPUs in a single node constrains the ability of \sysname to promote DoP values. However, a fixed static DoP is not optimal for all arrival rates and can not dynamically adjust in online serving. For example, at an arrival rate of 0.5, \sysname reduces p99 by up to 36.6\% and average latency by up to 44.4\% compared to a static DoP of 4, which is optimal at an arrival rate of 0.25.

{
\begin{figure*}[t]
\begin{center}
\centerline{\includegraphics[width=\textwidth]{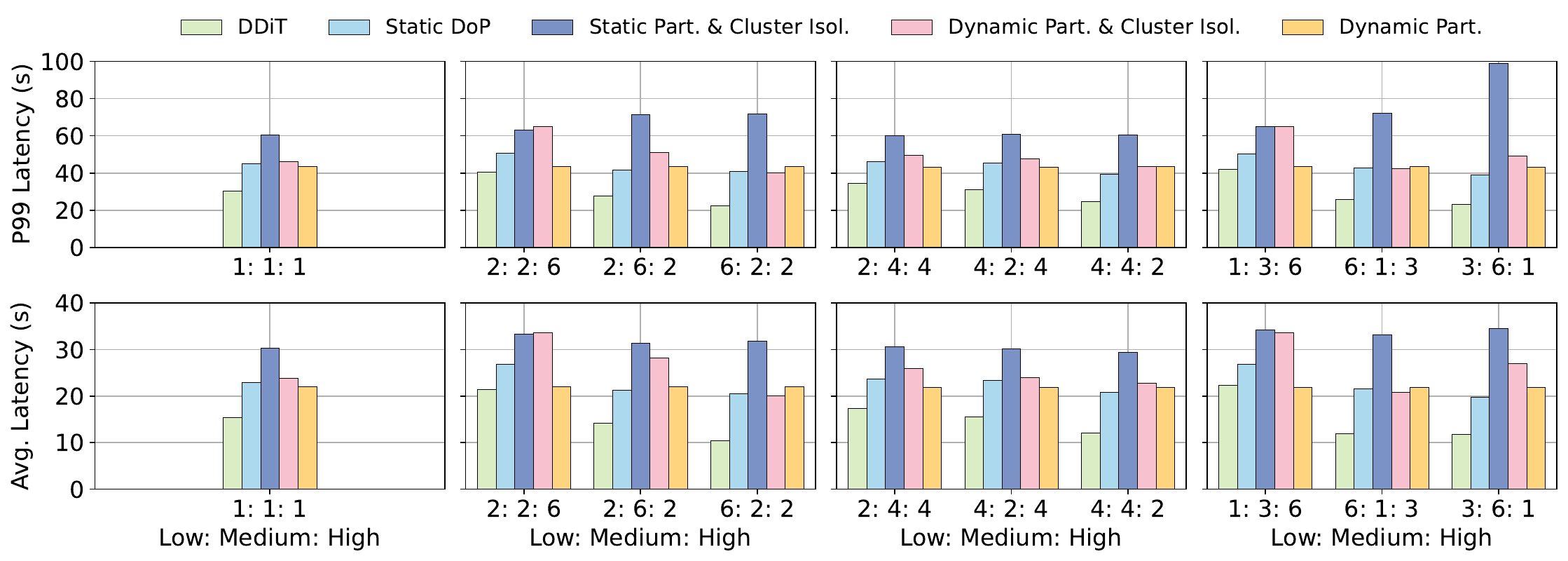}}
\caption{Multi-node end-to-end performance.}
\label{fig-mock}
\end{center}
\end{figure*}
}
{
\begin{figure}[t]
\begin{center}
\centerline{\includegraphics[width=0.48\textwidth]{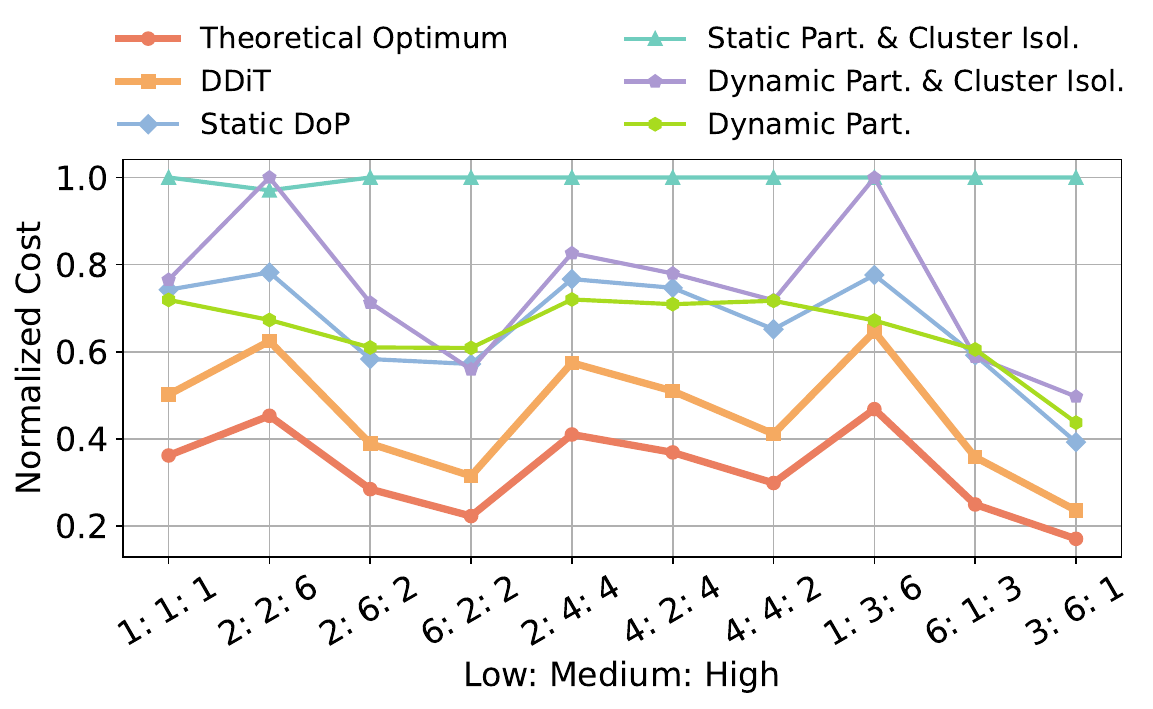}}
\captionsetup{belowskip=-0.8cm}

\caption{Multi-node monetary cost (Lower Cost indicates a better solution).}
\label{fig-mock-cost}
\end{center}
\end{figure}
}
\par For higher arrival rates (eg., 0.75 or 1.0), the optimal static DoP value remains 2. \sysname\ reduces p99 latency by up to 19.23\% and average latency by up to 29.5\% across different request distributions. As the arrival rate increases, the DiT-VAE decoupling and demotion launch mechanisms of \sysname effectively reduce resource contention among consecutive requests. Additionally, DoP promotion during execution ensures that each request maintains a reasonable execution time.
%
%
Finally, in the burst scenario, where requests arrive simultaneously and the system load is extremely high, a lower DoP enhances system-level concurrency under heavy load. As a result, setting the static DoP to 1 delivers the best performance. In this case, \sysname reduces p99 latency by up to 20.7\% and average latency by 21.7\%.
%

\par Both SPCI and DPCI cause resource wastage due to cluster isolation. As shown in Figure \ref{fig-total-comparison}, this inefficiency can worsen the performance as the arrival rate increases. Compared with SPCI, \sysname achieves a p99 latency reduction of approximately 27.6\% to 48.6\% across arrival rates from 0.25 to burst for different request resolution distributions. For DPCI, the reduction is around 14.5\% to 75.1\%. This difference arises because the DoP within each cluster in the dynamic partition is set based on the $B$ value, resulting in fewer model instances per cluster compared to the static partition. For DP, cross-cluster resource sharing completely compensates for the shortage of model instances. At lower arrival rates (0.25 and 0.5), \sysname achieves a 4.9\% reduction in p99 latency. This improvement increases to 22\%–24\% when the arrival rate reaches 1.0 or under burst conditions. The impact of \sysname on average latency follows the same trend as its effect on p99 latency compared to these two baselines.
{
\begin{figure}[t]
\begin{center}
\centerline{\includegraphics[width=0.48\textwidth]{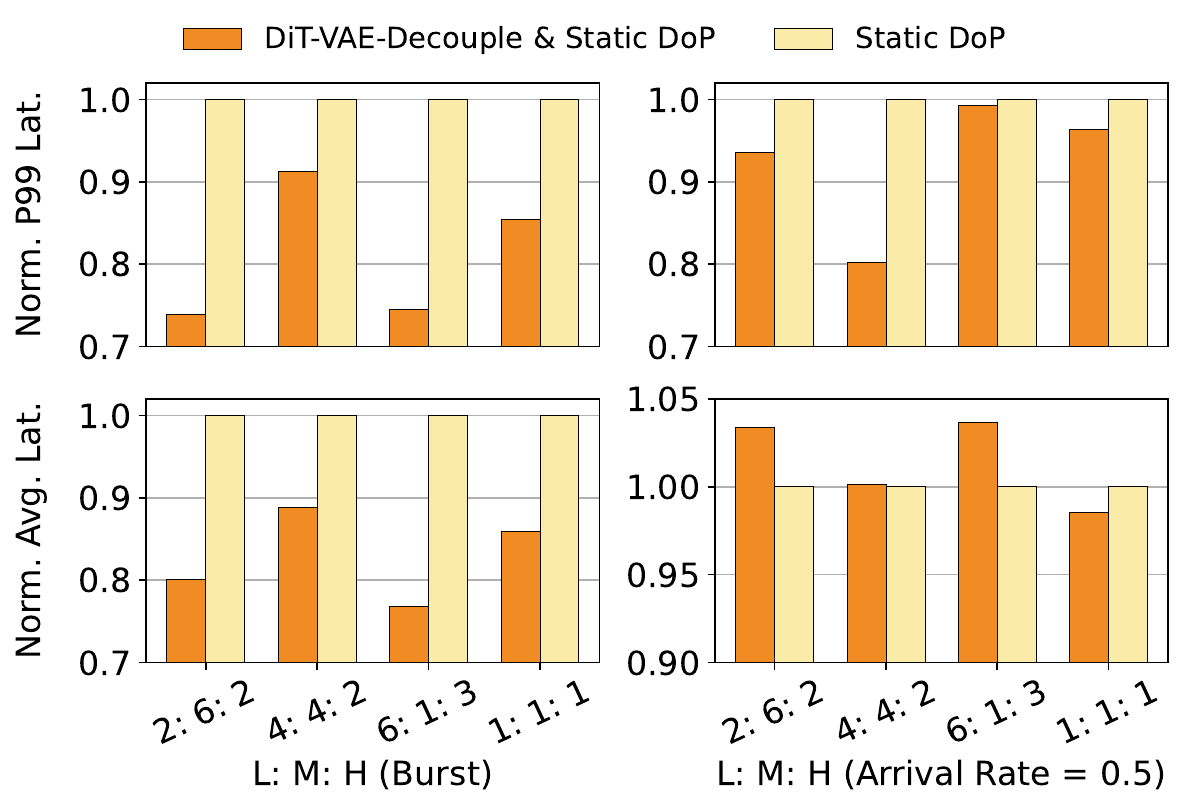}}
\captionsetup{belowskip=-0.8cm}

 \caption{Breakdown of DiT-VAE Decoupling Performance, including normalized P99 and Average Latency across Different Resolution Proportions (Lower Latency indicates Better Performance)}
\label{fig-bd-decouple}
\end{center}
\end{figure}
}

\noindent \textbf{Multi-Node Performance.} We evaluate the scalability of \sysname and the baselines in an emulated 8-node, 64-GPU cluster. Our single-node experiments reveal that \sysname becomes increasingly effective as the system load increases. Therefore, in our multi-node experiments, we focus on burst scenarios and use the monetary cost metric as described in \S\ref{sec-setup}. 

In Figure \ref{fig-mock}, we first present the average and p99 latencies with the static DoP fixed at 4. \sysname achieves at least a 30.4\% reduction in p99 latency and 30\% in average latency across four baselines. We observe that the SPCI method performs the worst, while the DPCI approach shows a slight improvement. This improvement is due to dynamic partitioning, which sets the DoP of each cluster based on the $B$ value, increasing concurrency for low-resolution requests. However, both SPCI and DPCI lead to resource wastage due to isolated clusters dedicated to different request types. Among the two methods without cluster isolation, DP demonstrates superior overall performance due to the suitable parallelism from DoP values.

\par Furthermore, we compare the monetary cost of \sysname with the four baselines across different request distributions. As shown in Figure \ref{fig-mock-cost}, \sysname reduces cost by up to 30.4\% relative to four baselines, using only 1.39× the theoretical optimum, whereas the best-performing baseline reaches 2.08× the optimum.

%
%

{
\begin{figure}[t]
\begin{center}
\centerline{\includegraphics[width=0.48\textwidth]{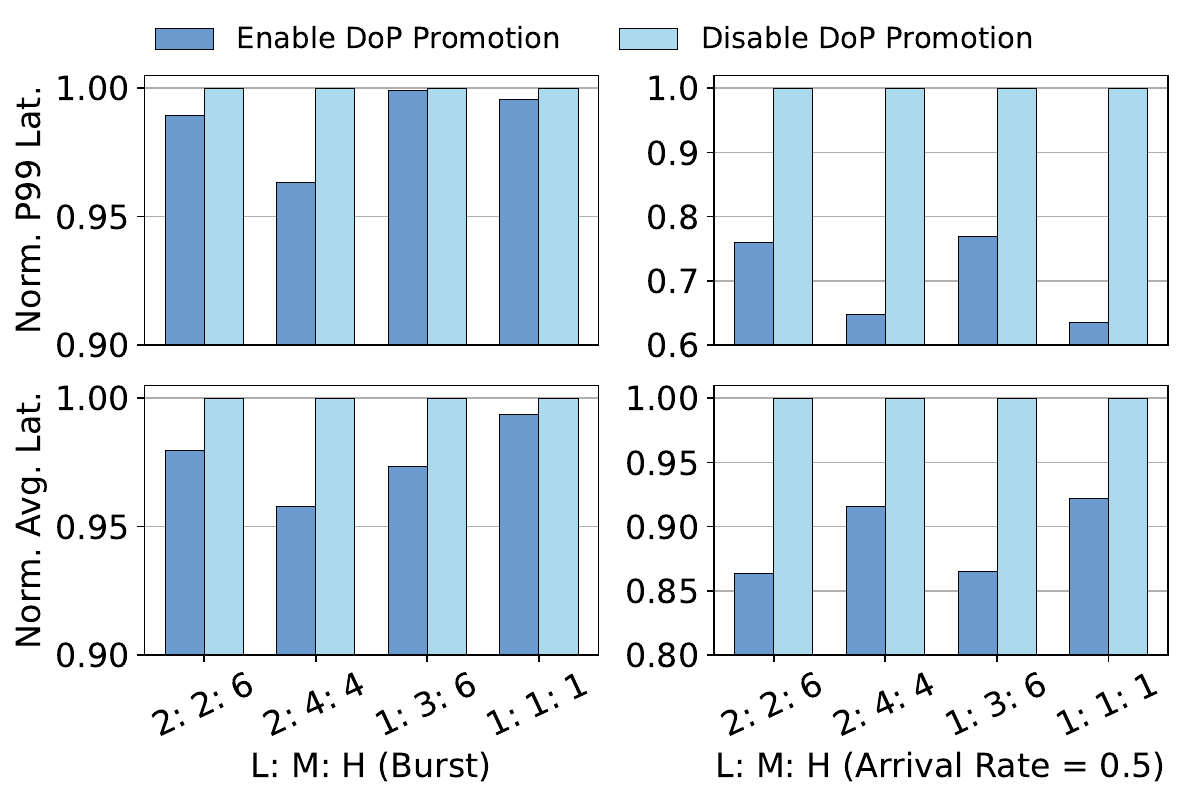}}

\captionsetup{belowskip=-0.8cm}
\caption{Breakdown DoP promotion.}
\label{fig-promote-decouple}
\end{center}
\end{figure}
}

\subsection{Breakdown}
\par \textbf{Effectiveness of DiT-VAE-Decouple.} We conduct an ablation study to evaluate the effectiveness of decoupling DiT-VAE by integrating this mechanism into static DoP. We set the DoP value to two on a single node in the experiment. As shown in Figure \ref{fig-bd-decouple}, decoupling DiT-VAE reduces the p99 latency by up to 20\%, while maintaining a similar average latency under an arrival rate of 0.5. We find that it can be attributed to the GPU release mechanism, where an odd number of GPUs are released after the DiT phase, while subsequent requests require an even number of GPUs in all scenarios except for one GPU. Under burst conditions, it improves the p99 latency by up to 26.1\% and the average latency by up to 23.2\% . This results in greater improvements under higher system loads, consistent with the end-to-end experiments.

\noindent \textbf{DoP Promotion.} To evaluate the effectiveness of DoP promotion in \sysname, we conduct an experiment by enabling and disabling it. As shown in Figure \ref{fig-promote-decouple}, DoP promotion improves the p99 latency by up to 35.2\% and the average latency by up to 13.5\%. However, it has minimal effect on accelerating requests under burst workload. We find that DoP promotion extends the waiting time for subsequent requests in an overutilized system, negating the benefits of acceleration. In contrast, it effectively leverages newly released resources in an underutilized system.

\noindent \textbf{Transfer \& Scale up Overhead.} We evaluate the overhead of transferring and scaling up from DoP promotion under different DoP values and request resolutions. As shown in Figure \ref{fig-overhead}, it introduces less than 1 millisecond of delay. Compared to the several seconds to tens of seconds required for DiT execution, these overheads are negligible.
 

{
\begin{figure}[t]
\begin{center}
\centerline{\includegraphics[width=0.48\textwidth]{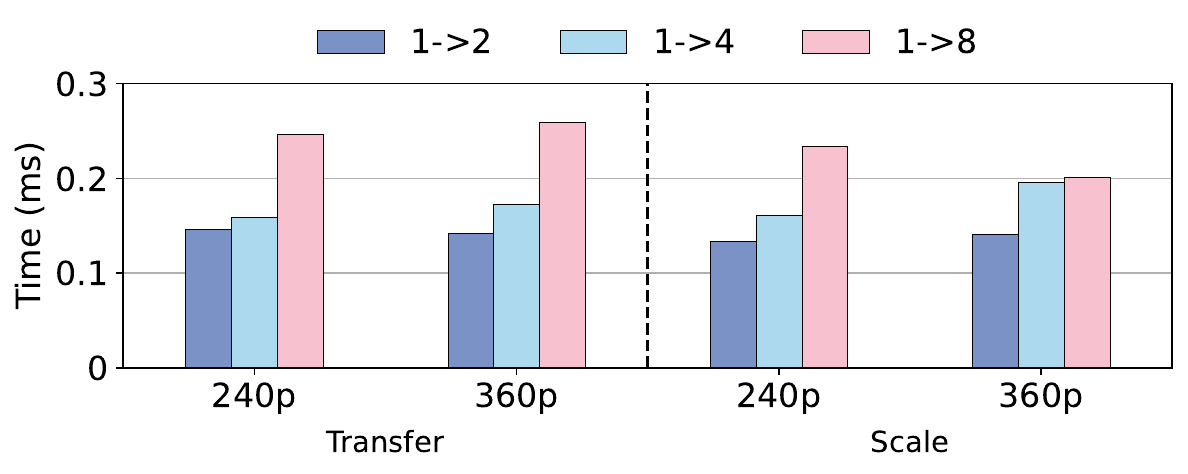}}
 \caption{Transfer \& scale up time overhead.}
\label{fig-overhead}
\end{center}
\end{figure}
}
\section{Related Work}
\noindent\textbf{Efficient Transformer Serving.} 
Existing transformer serving systems utilize efficient GPU operators for attention mechanisms, such as Flash Attention \cite{flashattention-nips22} and Flash-Decoding \cite{flashdecoding++-2023}, which are orthogonal to our work and have been integrated into \sysname. More recently, some works \cite{yu2022orca, vllm-sosp23} introduced batching techniques to enhance the system throughput. However, batching is incompatible with T2V generation due to its high computational demands.

\noindent\textbf{Sequence Parallelism.} 
SP is a specialized technique for distributing long sequences and activations across multiple devices, initially proposed to accelerate the training of long-context LLMs \cite{deepspeed-sc22, korthikanti2023reducing}. RingAttention \cite{li2021sequence} introduces a novel approach to partitioning the sequence dimension using a ring-style peer-to-peer (P2P) communication pattern, facilitating the transfer of keys and values across GPUs. Striped Attention \cite{brandon2023striped} enhances this by optimizing token distribution to achieve balanced attention computation. DSP \cite{zhao2024dsp} further advances the field by proposing an adaptive sequence parallelism abstraction for multi-dimensional transformers, dynamically switching the parallel dimension among sequences based on the computation stage. However, the serving system must determine the DoP of sequence parallelism before model deployment, which poses challenges for handling sequences of varying lengths efficiently.

\noindent\textbf{Caching}. Caching reuses the intermediate results during the inference to reduce computation, optimize GPU utilization, and lower latency without additional training. Unlike caching in LLMs, which reuses the same data without sacrificing accuracy, T2V leverages similarity features between data for efficiency. DeepCache \cite{ma2024deepcache} utilizes high-level convolutional features in consecutive denoising steps. PAB \cite{pab-2024}, FORD \cite{selvaraju2024fora} and $\Delta$\text{-DiT} \cite{chen2024delta} extend a similar strategy in the attention mechanism. Some works utilize caching to overlap data transfer and computation, enabling efficient distributed inference \cite{pipefusion-2024, distrifusion-2024}. However, leveraging the similarity between outputs of adjacent steps may compromise the quality of the generated images or videos.

\section{Conclusion}
To efficiently serve DiT-based T2V models, \sysname\ proposes DiT-VAE decoupling, DoP promotion, and a complementary efficient scheduling algorithm. To further validate the effectiveness of \sysname, we also propose an optimal scheduling algorithm under ideal conditions. Our experiments show that \sysname can significantly reduce both p99 and average latencies for T2V tasks across various workloads.

\bibliographystyle{plain}
\bibliography{paper}

\twocolumn

\appendix

\section{Queue Model} \label{appendix-a}
We clarify the queue model in Algorithm \ref{Theoretical Upper Bound Solver} here. If the time interval between task arrivals follows a Poisson distribution. Given that the number of model instances (servers) $\alpha \ge 1$, the arrival rate $\lambda$ and the execution time (service time) $d$ is fixed for task type ${j}$, and assuming the utilization ratio $\rho<1$, we model the scenario using either the M/D/1 queue or M/D/c queue \cite{queue_theory}, as appropriate. The average resource occupancy time for a single GPU (\textbf{Occupy}(...)), which is also the average time spent in a queue system, is directly computed as it follows the M/D/1 ($\alpha=1$) queue:
    \begin{equation}
    W_{M/D/1}(type_{j})=\frac{1}{\mu}+\frac{\rho}{2\mu(1-\rho)}
    \label{md1}
    \end{equation}
where $\mu=\frac{1}{d}$ and $\rho=\frac{\lambda\cdot x_{j}}{\mu}$. Conversely, computing the average resource occupancy time for the M/D/c ($\alpha>1$) queue is highly complex. Thus, we approximate it using M/M/c queue by Eq. \ref{mdc1} from \cite{mdc_assump,queue_theory}, where $\mu=\frac{1}{d}$, $\rho=\frac{\lambda\cdot x_{j}}{\alpha\mu}$, $r=\frac{\lambda\cdot x_{j}}{\mu}$ and $p_{0}=(\frac{r^{\alpha }}{\alpha !(1-\rho)}+\sum_{s=0}^{\alpha -1}\frac{r^s}{s!})^{-1}$. To further reduce time complexity, we employ Stirling's formula \cite{factorial_assump}, $n!\approx \sqrt{2\pi n}(\frac{n}{e})^n$, to efficiently compute integer factorials as follows: 
    \begin{equation}
    W_{M/D/c}(type_{j})\approx \frac{W_{M/M/c}(type_{j})}{2}=\frac{\frac{1}{\mu}+(\frac{r^\alpha }{\alpha !(\alpha \mu)(1-\rho)^2})p_{0}}{2}
    \label{mdc1}
    \end{equation}

\section{Scheduling Algorithm of DDiT} \label{sec:appendix-b} 

\begin{algorithm}
\footnotesize
\caption{Greedy Scheduling Algorithm of \sysname}
\begin{algorithmic}[1]
\State $GS \gets Global\_Scheduler,\ GS_{G} \gets Free\_GPU\_List$
\State $GS_{PT}\gets Promote\_Table,\ GS_{NG} \gets New\_GPU\_Event$
\State $RQ \gets Running\_Queue,\ WQ\gets Waiting\_Queue$
\Function{Global\_Schedule}{$GS$}
\While{$True$}
\If{$GS_{NG}.is\_set()$} 
\State $\underset{r\in GS_{PT}}{r_{starv.}}.update(r_{\textbf{Meta.}})$
\State $\textbf{Desc.\_Sort}(GS_{PT},\ key=r_{starv.})$
\State $r_{last\_step}\gets r_{cur\_step}$

\For{$r\in GS_{PT}$}
\State $r_{GPU\_ids}\underset{\textbf{async}}{\gets}\textbf{Try\_Best\_Alloc.}(r,\ GS_{G},\ r_{GPU\_ids})$
\If{$\textbf{len}(r_{GPU\_ids})==r_{opt\_GPU\_num}$}
\State $GS_{PT}.remove(r)$
\EndIf
\EndFor
\State $GS_{NG}.clear()$
\EndIf

\For{$r \in WQ$}
\State $r_{GPU\_ids}\gets \textbf{Try\_Best\_Alloc.}(r,\ GS_{G},\ \emptyset)$
\If{$r_{GPU\_ids}\ne \emptyset$}
\State $W.remove(r)\ \&\&\ R.add(r)$
\If{$\textbf{len}(r_{GPU\_ids})\ne r_{opt\_GPU\_num}$}
\State $GS_{PT}.add()\underset{\textbf{async}}{\gets}r$
\EndIf
\EndIf
\EndFor

\EndWhile
\EndFunction

\end{algorithmic}
\label{ddit scheduler 1}
\end{algorithm}

We present the details of how \sysname schedules greedily in Algorithm \ref{ddit scheduler 1}. We first define running, hungry, and waiting as the statuses of requests in \sysname. Hungry refers to a running request whose allocated GPUs are fewer than the optimal number. We use a running queue ($RQ$) or waiting queue ($WQ$) to place the corresponding request, while the hungry request is stored in the priority table $GS_{PT}$. In \sysname, the hungry status is given higher priority than waiting requests, considering the arrival time. Therefore, the core idea of our scheduling policy is to supply additional GPUs to hungry requests as needed while promptly allocating GPUs to new incoming requests. 

\par The new incoming request is initially set to a waiting status and placed in $WQ$. The global scheduler manages the scheduling of the request and communicates with the resource allocator for GPU allocation. As described in lines 15 to 20 of Algorithm \ref{ddit scheduler 1}, the process starts from the optimal GPU count and decreases incrementally to ensure timely execution. If the allocated GPU count matches the optimal value, the request is directly inserted into $RQ$ and sent to the engine unit with a control message. Otherwise, an additional step is taken to mark the request as hungry and place it in $GS_{PT}$.

\par We continuously check $GS_{PT}$ for hungry requests to acquire available resources. To prioritize DoP promotion, we define the starvation time of requests, giving priority to those with GPU allocations below the optimal number in e.q. \ref{eq-hug-priority}, therefore, the $r_{\textbf{Meta.}}$ in line 7 refers to $r_{cur\_step}$, $r_{last\_step}$, $r_{cur\_step\_time}$ and $r_{opt\_step\_time}$. The $GS_{NG}$ is triggered by the step \textcircled{4} and \textcircled{6} in Figure \ref{fig-lifecycle-req}.

\section{Single-Node Performance II} \label{sec:appendix-e2e}
\begin{figure*}[!t]
\centering
\includegraphics[width=\textwidth]{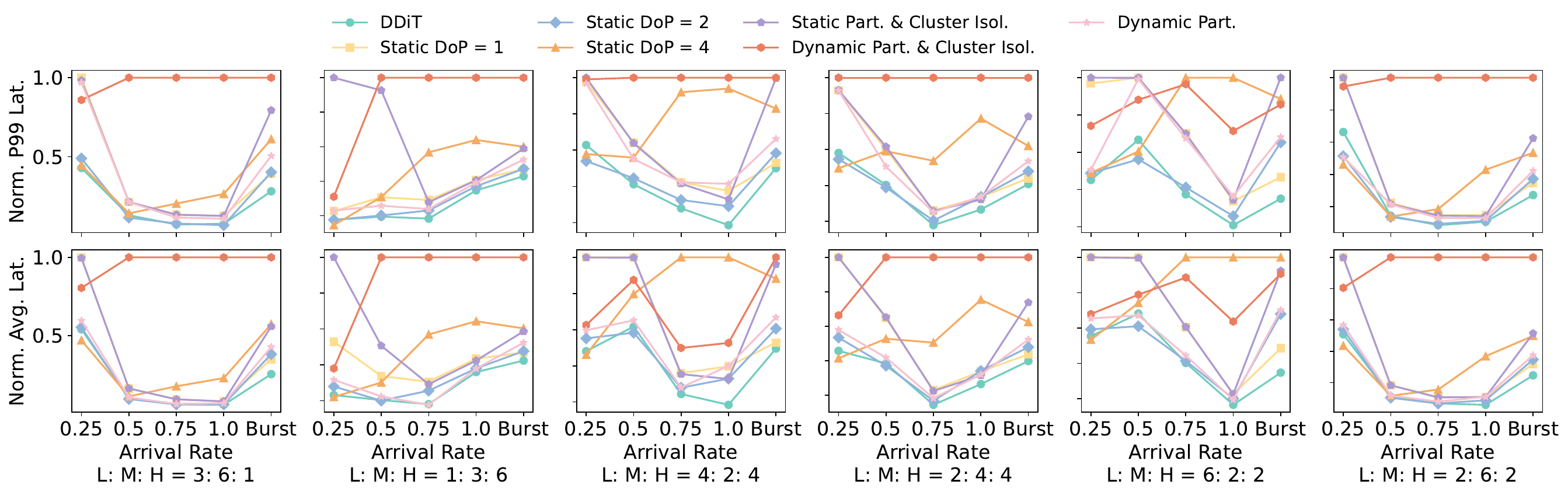}
\caption{Single-node end-to-end performance II.}
\label{fig-total-comparison2}
\end{figure*}

This supplement to the single-node experiment in our evaluation reveals a trend consistent with our earlier observations: the higher the arrival rate, the more effective \sysname becomes. As shown in Figure \ref{fig-total-comparison2}, across these resolution distributions, \sysname achieves an average reduction of up to 30\% in both p99 and average latencies.

\begin{figure*}[!t]
\centering
\includegraphics[width=\textwidth]{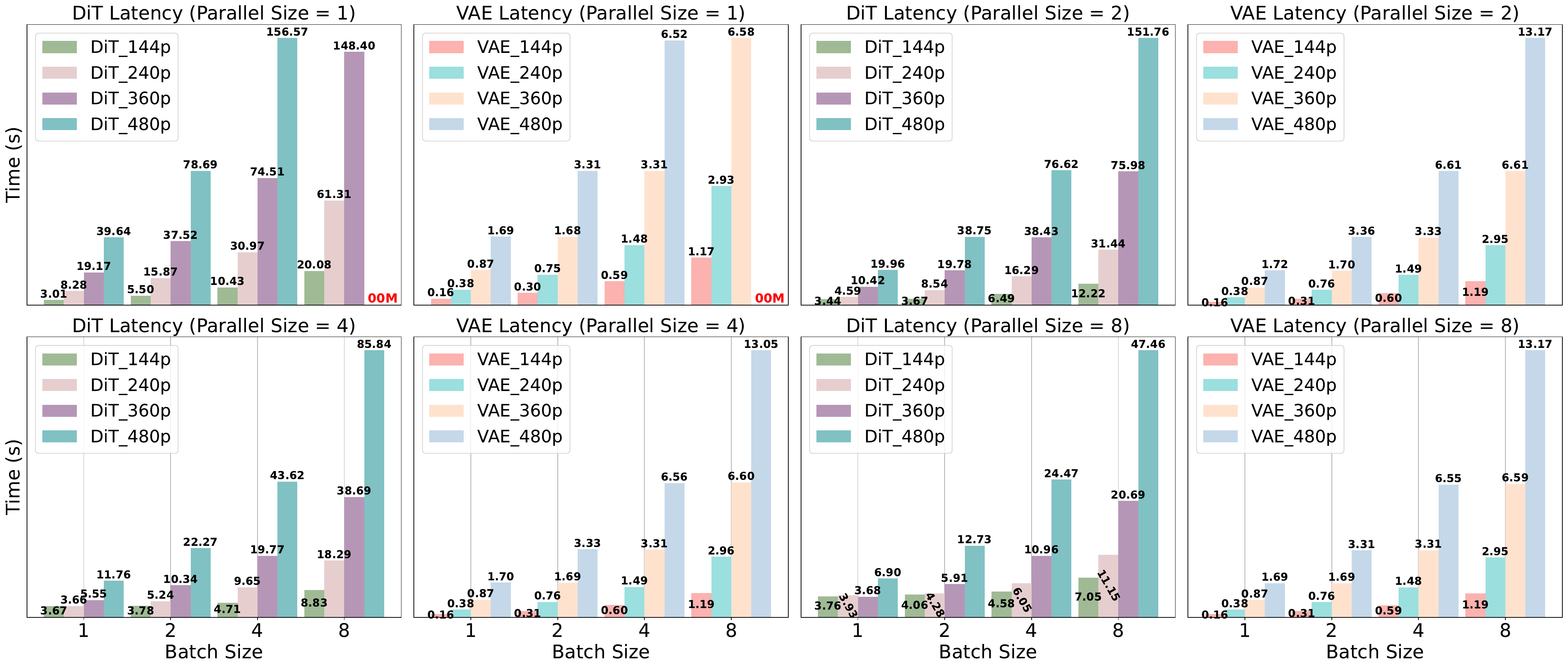}
\caption{Characterization of DiT and VAE under various DoPs and batch sizes.}
\label{fig-total-charac}
\end{figure*}

\end{document}